\documentstyle[12pt,fleqn,epsfig]{article}

\begin{document}

\begin{titlepage}

\begin{center}
\textbf{\large Short-range correlations and neutrinoless 
double beta decay}
\vspace{8mm}

\centerline{M. Kortelainen$^a$, O. Civitarese$^b$, J. Suhonen$^a$, 
J. Toivanen$^a$}

\vspace{2mm}

\textit{ $a)$ Department of Physics, University of Jyv\"askyl\"a, 
P.O.Box 35, FIN-40351, Jyv\"askyl\"a, Finland}

\vspace{2mm}

\textit{ $b)$ Department of Physics, University of La Plata, c.c.
67, (1900) La Plata, Argentina. }

\end{center}

\vspace{4mm}
\centerline \textbf{ \today }

\vspace{4mm}

\begin{abstract}{
In this work we report on the effects of short-range correlations
upon the matrix elements of neutrinoless double beta decay
($0\nu\beta\beta$). We focus on the calculation of the matrix
elements of the neutrino-mass mode of $0\nu\beta\beta$ decays of
$^{48}$Ca and $^{76}$Ge. The nuclear-structure components of the
calculation, that is the participant nuclear wave functions, have
been calculated in the shell-model scheme for $^{48}$Ca and in the
proton-neutron quasiparticle random-phase approximation
(pnQRPA) scheme for $^{76}$Ge. We compare the traditional
approach of using the Jastrow correlation function with the more
complete scheme of the unitary correlation operator method (UCOM). Our
results indicate that the Jastrow method vastly exaggerates the
effects of short-range correlations on the $0\nu\beta\beta$ nuclear
matrix elements.}
\end{abstract}

\vspace{0.5cm}

\noindent
PACS number(s): 21.60.Cs, 23.40.Hc, 27.40.+z.

\vspace{0.3cm}


\bigskip
\noindent
To appear in \emph{Physics Letters B (2007)}

\end{titlepage}

A good knowledge of the nuclear matrix elements governing the
decay rates of neutrinoless double beta ($0\nu\beta\beta$) decay
is mandatory if one wants to extract information about the
neutrino mass from the current experimental limits for the
half-lives of $0\nu\beta\beta$-decay transitions
\cite{exp1,compil}. The standard theoretical methods which are
suitable for the calculation of the relevant nuclear matrix
elements can be found in the literature, e.g. in \cite{teo1,teo2}.
Although many of the formalisms are well established, difficulties
arise from the approximations which are needed in order to perform
the actual calculations.

In the mass mode of the
$0\nu\beta\beta$ decay the involved two nucleons exchange a light Majorana
neutrino \cite{DOI85}. Average value of the
exchanged momentum is of the order of
$100-200\,\textrm{MeV/c}$ and thus the involved nucleons are on average at
close distance from each other. There is, however, a minimum relative
distance of the order of $1\,\textrm{fm}$ after which the two
nucleons may eventually overlap. In nuclear matter this overlapping cannot
happen and in theoretical description of the $0\nu\beta\beta$ decay
one needs to take into account this fact. Based on this it has been
argued \cite{miller} that special measures have to be taken when performing
nuclear-structure calculations using the mean-field picture with
residual two-body interactions between the two interacting
nucleons. In the case of the $0\nu\beta\beta$ decay these measures boil
down to introducing an explicit Jastrow type of correlation function into the
involved two-body transition matrix elements \cite{teo1}. Using this
method in the numerical calculations of $0\nu\beta\beta$-decay matrix 
elements considerable corrections to the involved Fermi and
Gamow--Teller nuclear matrix elements were reported \cite{teo1,tueb}. 

In \cite{KRM92} a different method
was used to explicitly take into account the short-range
correlations. This approach is based on the use of the Horie--Sasaki
method to evaluate the involved radial form factors and the short-range
correlations were considered to arise from the $\omega$ exchange in
the nucleon-nucleon interaction \cite{BRO77,TOW87}. Contrary to
\cite{teo1,tueb}, in this approach only relatively small corrections 
to the involved nuclear matrix elements were obtained \cite{KRM92,BAR99}.
Instead of using the above described methods, one can
use the more complete new concept of unitary correlation operator 
method (UCOM) \cite{feldmeier98} to take into account
the short-range effects in $0\nu\beta\beta$ decay. In this method a
unitary correlation operator moves a pair of nucleons away from each
other whenever they start to overlap. This method also conserves the
probability normalization of the relative wave function.

In this work we address the important issue of short-range
correlations in the computation of nuclear matrix elements involved in
the neutrinoless double beta decay. We have used both the UCOM
and Jastrow methods and we compare the results for the
ground-state-to-ground-state $0\nu\beta\beta$ decays of $^{48}$Ca
and $^{76}$Ge. For $^{48}$Ca we calculate the relevant nuclear wave
functions in the solid theoretical framework of the nuclear
shell model. In order to accomplish
this we have used the OXBASH code \cite{shell}, which is actually
available to any practitioner in the field. For $^{76}$Ge we have used
the framework of the proton-neutron quasiparticle random-phase approximation
(pnQRPA), suitable for calculations of nuclear properties of
medium-heavy and heavy nuclei.

Our calculated results show that the reduction caused by the
inclusion of the short range correlations depends on the multipole
which contributes to the $0\nu\beta\beta$ matrix element. In
addition, the strength distributions of the multipoles
are practically unaltered by the short-range correlations, suggesting
the effect to be just an overall multipole-dependent scaling.
This scaling factor is close to unity for all multipoles in the UCOM
scheme, but acquires strongly reduced values for high multipoles in 
case of the Jastrow method.

We start the quantitative scrutiny of the effects of short-range 
correlations by briefly presenting the central ideas behind our computations.
By assuming the neutrino mass mechanism to be the dominant one
in the $0\nu\beta\beta$ decay one can write as a good approximation
the inverse of the half-life as \cite{teo2}
\begin{equation} \label{eq:0nbbhl}
\left[ t_{1/2}^{(0\nu)}\right]^{-1} = G_{1}^{(0\nu)}
\left( \frac{\langle m_{\nu}\rangle }{m_{\rm e}} \right)^{2}
\left( M_{\rm GT}^{(0\nu)} - \left( \frac{g_{\rm V}}{g_{\rm A}}
\right)^{2} M_{\rm F}^{(0\nu)}\right)^{2}\, ,
\end{equation}
where $m_{\rm e}$ is the mass of the electron and
\begin{equation}
\langle m_{\nu}\rangle = \sum_{j} \lambda^{\rm CP}_{j} m_{j}
|U_{{\rm e}j}|^{2}
\end{equation}
is the effective mass of the neutrino, $\lambda^{\rm CP}_{j}$
being the CP phase. Furthermore, the quantity  $G_{1}^{(0\nu)}$
of Eq. (\ref{eq:0nbbhl}) is the leptonic phase-space factor
defined in \cite{teo2}. The double Gamow-Teller and double Fermi
nuclear matrix-elements in (\ref{eq:0nbbhl}) are defined as \cite{DOI85} 
\begin{eqnarray}
M_{\rm F}^{(0\nu)} & = & \sum_{a} (0^{+}_{f} || h_{+}
(r_{mn},E_{a}) || 0^{+}_{i}) \, , \\
M_{\rm GT}^{(0\nu)} & = & \sum_{a} (0^{+}_{f} || h_{+}
(r_{mn},E_{a}) ( \sigma_{m}\cdot \sigma_{n} ) || 0^{+}_{i}) \, .
\end{eqnarray}
Here the summation runs over all states $a$ of the intermediate
nucleus, which in this case are $^{48}$Sc and $^{76}$As.
The definition of the neutrino potential $h_{+}(r_{mn},E_{a})$
can be found in Refs. \cite{teo1,teo2,DOI85}. 

The traditional way \cite{teo1} to include the short-range 
correlations in the $0\nu\beta\beta$ matrix elements is 
by introducing the Jastrow correlation function $f_{\rm J}(r)$. 
It has to be noted that this particular variant of the Jastrow
function is a rudimentary one and does not do full justice to the
name Jastrow correlations. For example, in light nuclei accurate
Monte-Carlo calculations are based on Jatrow-like
correlations which are variationally determined and have different
ansatz functions in different isospin channels. This fact notwithstanding
we choose to call here the rudimentary approach of \cite{teo1} as
Jastrow method since this is the term adopted by the 
double-beta-decay community.

The Jastrow
function depends on the relative distance $r=|{\bf r}_{1}-{\bf r}_{2}|$ 
of two nucleons, and in the Jastrow scheme one replaces the bare
$0\nu\beta\beta$ operator ${\mathcal O}$ by a correlated operator 
${\mathcal O}_{\rm J}$ by the simple procedure
\begin{equation} \label{eq:Osrc}
(0^{+}_{f} || {\mathcal O} || 0^{+}_{i}) \to
(0^{+}_{f} || {\mathcal O}_{\rm J} || 0^{+}_{i}) =
(0^{+}_{f} || f_{\rm J} {\mathcal O} f_{\rm J} || 0^{+}_{i}) \, .
\end{equation}
A typical choice for the function $f_{\rm J}$ is
\begin{equation} \label{eq:jastrow}
f_{\rm J}(r) = 1 - e^{-ar}\left( 1- br^{2} \right) \, ,
\end{equation}
with $a=1.1 \, {\rm fm}^{2}$ and $b=0.68 \, {\rm fm}^{2}$.
As a result, the Jastrow function effectively cuts out the 
small $r$ part from the relative wave function of the two nucleons.
For this reason, the traditionally adopted Jastrow procedure 
does not conserve the norm of the relative wave function and 
one should use, in principle, the operator
\begin{equation}
  {\mathcal O}_{\rm J} = \frac{f_{\rm J}{\mathcal O} f_{\rm J}}
{\langle \Psi \vert f_{\rm J}f_{\rm J} \vert \Psi \rangle} 
\end{equation}
when the initial and final states are the same, here denoted 
by $\vert \Psi \rangle$. For different initial and final states 
normalization is even more problematic. Even a proper normalization
does not guarantee that the short-range correlations would be
correctly treated by the Jastrow procedure.

To circumvent the difficulties associated with the use of a Jastrow
function one can adopt the more refined unitary correlation operator 
method (UCOM) \cite{feldmeier98} when discussing short-range effects 
in the $0\nu\beta\beta$ decay. In the UCOM one obtains the 
correlated many-particle state $\vert \tilde{\Psi} \rangle$ from 
the uncorrelated one as
\begin{equation}
\vert \tilde{\Psi} \rangle = C \vert\Psi\rangle \, ,
\end{equation}
where $C$ is the unitary correlation operator. The operator $C$ 
is a product of two unitary operators: $C=C_{\Omega}C_{r}$, where 
$C_{\Omega}$ describes short-range tensor correlations and 
$C_{r}$ central correlations. Due to unitarity of the operator 
$C$, the norm of the correlated state is conserved. Moreover,
one finds for the matrix element of an operator $A$ 
\begin{equation}
\langle \tilde{\Psi} \vert A \vert \tilde{\Psi}' \rangle
= \langle \Psi \vert C^{\dag}A C \vert \Psi' \rangle
= \langle \Psi \vert \tilde{A}  \vert \Psi' \rangle \ ,
\end{equation}
so that it is equivalent to use either correlated states
or correlated operators.

The exact form
of the operator $C$ is gained by finding the minimum of the
Hamiltonian matrix element
$\langle \Psi\vert C^{\dag}HC\vert \Psi \rangle$.
Therefore, the choice of the two-body interaction in $H$ affects
also the form of $C$. Explicit expressions for the operators $C_{r}$ 
and $C_{\Omega}$ can be found in Refs. \cite{feldmeier98,roth05}. 
Application of these expressions to the double Gamow-Teller and Fermi 
matrix elements shows that the tensor correlations of $C_{\Omega}$ 
vanish and we are left with only the central correlations.

For $^{48}$Ca the nuclear-structure calculations were handled
by the shell-model code OXBASH \cite{shell}. In our calculations
we have used the FPBP two-body interaction of \cite{force}, which
was obtained by fitting the Kuo-Brown interaction to experimental
data. Due to the fact that we have limited our model space to the pf shell,
the $0\nu\beta\beta$ matrix elements are composed of only 
positive-parity states. The shell-model calculations had to be
truncated by requiring that the minimum number of particles in 
the 0f$_{7/2}$ orbital be 4.

\begin{table}[htb]
\caption{Multipole decomposition and the total value of the matrix element
$M_{\rm GT}^{(0\nu)}$ for $^{48}$Ca. The cases are: no short-range 
correlations included (bare), 
with Jastrow correlations and with UCOM correlations using Bonn-A and 
Argonne V18 parametrizations.}
\label{t:gtcon}
\begin{center}\begin{tabular}{lrrrr}
\hline
 & & & \multicolumn{2}{c}{UCOM} \\
$J^{\pi}$ & bare & Jastrow & Bonn-A & AV18 \\
\hline
$1^{+}$ & -0.330 & -0.305 & -0.322 & -0.319 \\
$2^{+}$ & -0.117 & -0.092 & -0.108 & -0.104 \\
$3^{+}$ & -0.327 & -0.246 & -0.302 & -0.293 \\
$4^{+}$ & -0.066 & -0.035 & -0.054 & -0.051 \\
$5^{+}$ & -0.246 & -0.121 & -0.212 & -0.199 \\
$6^{+}$ & -0.042 & -0.008 & -0.030 & -0.027 \\
$7^{+}$ & -0.150 & -0.029 & -0.120 & -0.107 \\
sum     & -1.278 & -0.835 & -1.150 & -1.101 \\
\hline
\end{tabular}\end{center}\end{table}

\begin{table}[htb]
\caption{The same as Table \ref{t:gtcon} but for $M_{\rm F}^{(0\nu)}$.}
\label{t:fcon}
\begin{center}\begin{tabular}{lrrrr}
\hline
 & & & \multicolumn{2}{c}{UCOM} \\
$J^{\pi}$ & bare & Jastrow & Bonn-A & AV18 \\
\hline
$1^{+}$ & 0.000 & 0.000 &  0.000 &  0.000\\
$2^{+}$ & 0.185 & 0.145 &  0.174 &  0.169\\
$3^{+}$ & 0.000 & 0.000 & -0.001 & -0.001\\
$4^{+}$ & 0.116 & 0.061 &  0.102 &  0.096\\
$5^{+}$ & 0.000 & 0.000 & -0.002 & -0.002\\
$6^{+}$ & 0.061 & 0.012 &  0.050 &  0.045\\
$7^{+}$ & 0.000 & 0.000 & -0.002 & -0.002\\
sum     & 0.367 & 0.221 &  0.324 &  0.308\\
\hline
\end{tabular}\end{center}\end{table}

Our main results for $^{48}$Ca are presented in Tables \ref{t:gtcon} and
\ref{t:fcon}. In these tables we list the calculated 
multipole decomposition and total values of the matrix elements 
$M_{\rm GT}^{(0\nu)}$ and $M_{\rm F}^{(0\nu)}$ for four
different cases. In the first case, which we refer to as bare
matrix elements, we have not taken into account any short-range correlations.
In the second case the short-range effects were handled by the use of
the Jastrow function (\ref{eq:jastrow}) and the replacement
(\ref{eq:Osrc}). In the third and fourth cases we have used the
UCOM to account for the short-range effects. The Kuo-Brown interaction
was not derived via UCOM, as it should be if it were to be used in the
same calculation as the UCOM-derived double-beta operator.
To access the magnitude of the resulting effect, we have adopted
two different UCOM parameter sets in the present calculation. These two parameter 
sets were obtained by minimizing the energy for the Bonn-A and 
Argonne V18 potentials. Both of the used UCOM parameter sets 
can be found in \cite{neff03}.

As the results in Tables \ref{t:gtcon} and \ref{t:fcon} indicate, the 
differences between the results obtained by the use of the two UCOM 
parameter sets are small. Therefore, we expect that the results
obtained by the use of the Kuo-Brown UCOM parameters do not 
deviate significantly from the Bonn-A or Argonne V18 results.
We also note that there exist a small UCOM contribution to the double
Fermi matrix element $M_{\rm F}^{(0\nu)}$ coming from the odd-$J$ 
intermediate states. This is explained by the fact that in Ref.
\cite{neff03} slightly different parameters were given to the $S=0$ 
and $S=1$ channels.

\begin{figure}[htb]
\includegraphics[width=12cm]{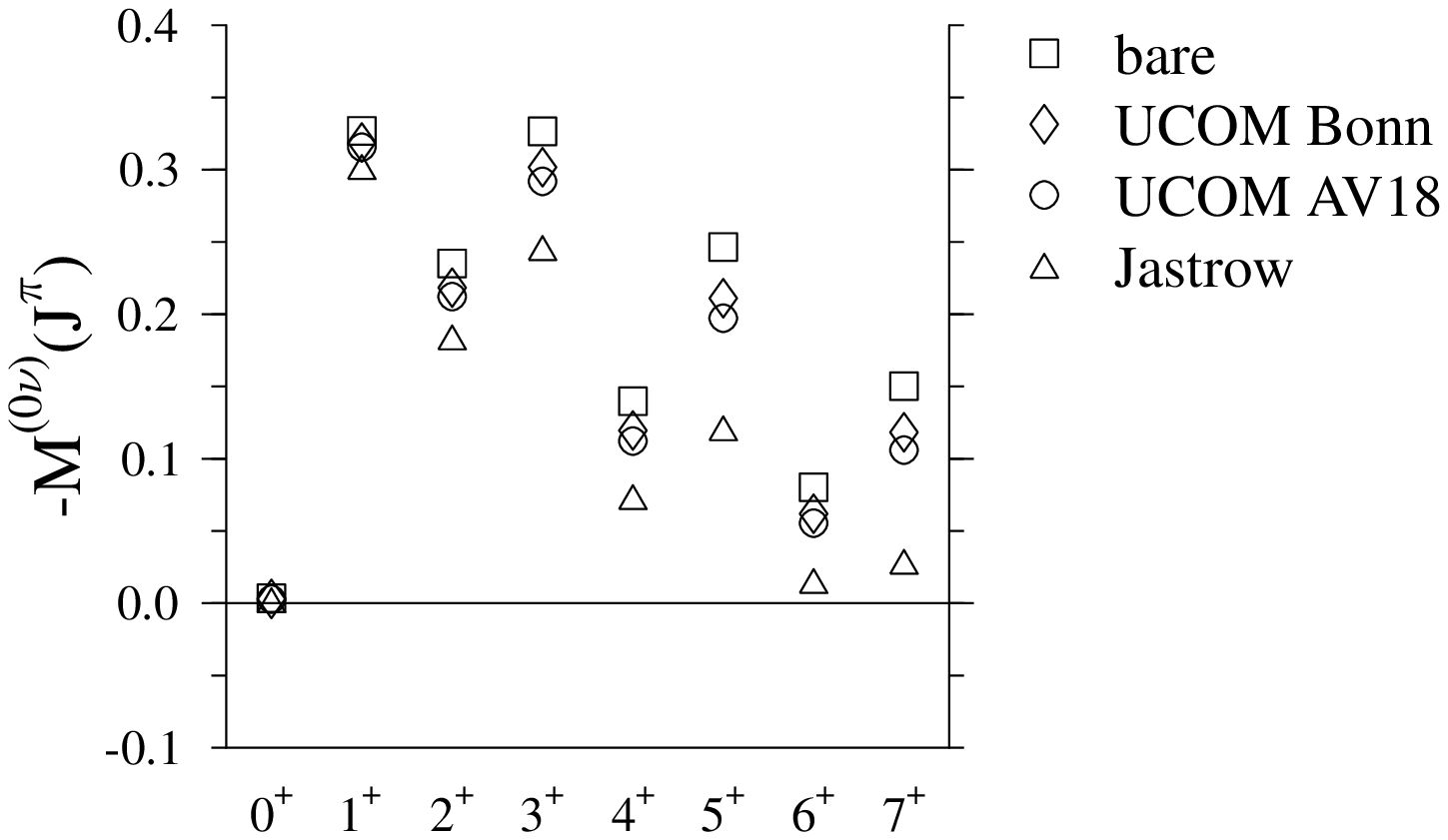}
\caption{Multipole decomposition of the total $0\nu\beta\beta$ decay 
matrix element
$M_{\rm GT}^{(0\nu)}-\left(g_{\rm V}/g_{\rm A}\right)^2 
M_{\rm F}^{(0\nu)}$ for $^{48}$Ca. The cases are: no short-range 
correlations included (bare),
with Jastrow correlations and with UCOM correlations using
the Bonn-A and Argonne V18 parametrizations.}
\label{f:totcomp}
\end{figure}

In Fig.~\ref{f:totcomp} we show graphically the multipole
decomposition of the total matrix element 
$M_{\rm GT}^{(0\nu)}-\left(g_{\rm V}/g_{\rm A}\right)^2 
M_{\rm F}^{(0\nu)}$ of (\ref{eq:0nbbhl})
for the four different cases of Tables
\ref{t:gtcon} and \ref{t:fcon}. The ratio $g_{\rm A}/g_{\rm V}=-1.254$ 
was used in this plot. As can be seen, the results obtained by using
the two different UCOM parameter sets do not differ significantly.
Also, one can see that the effects of the Jastrow or UCOM correlations
grow with increasing $J$ of the intermediate states. For the extreme 
case of the $7^{+}$ contributions the switching on of the 
Jastrow correlations changes the value of the matrix element 
$M_{\rm GT}^{(0\nu)}(7^{+})$ from $-0.150$ to $-0.029$, roughly
corresponding to a factor of $5$ reduction. At the same time 
the UCOM correlations produce only a 20\% -- 30\% reduction from the 
bare matrix element. It seems that in a situation like this
blind use of Jastrow correlations cuts out relevant parts of the
nuclear many-body wave function. From the tables one deduces that
the Jastrow correlations cause some 35\% -- 40\% reduction 
to the magnitudes of the total matrix elements, 
whereas the UCOM causes a reduction of 10\% -- 16\%. It is worth 
pointing out that our numbers for the Jastrow 
case coincide with the numbers of the corresponding earlier
calculation performed by the Strasbourg group \cite{RET95}.

\begin{figure}[htb]
\includegraphics[width=12cm]{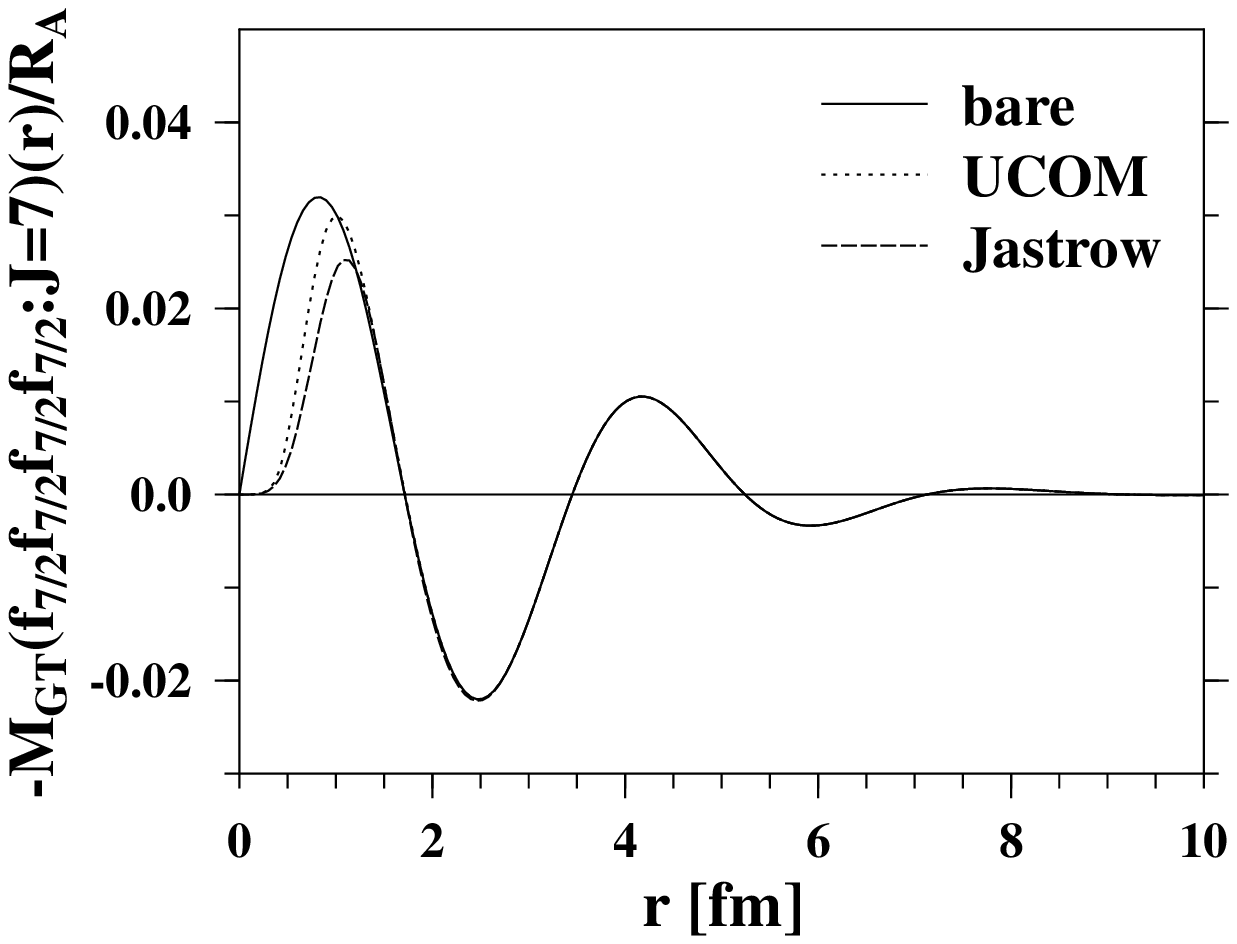}
\caption{Radial dependence of the two-particle 
Gamow-Teller $0\nu\beta\beta$ matrix element 
for $p=p'=n=n'=0f_{7/2}$ and $J=7$ in the case of $^{48}$Ca decay. 
Shown are the bare matrix element and Jastrow and
UCOM correlated matrix elements.}
\label{f:radme}
\end{figure}

To trace the source of differences between the Jastrow and UCOM
corrected matrix elements we show for the $^{48}$Ca decay in 
Fig.~\ref{f:radme} the radial dependence of the two-particle 
Gamow-Teller $0\nu\beta\beta$ matrix element 
in the special case of $p=p'=n=n'=0f_{7/2}$ and $J=7$ (this is
the contribution to Eq.~(4.16) of \cite{teo1} without including
the one-body transition densities and the overlap of the two complete
sets of pnQRPA states). 
The oscillator parameter value $b=2.0\,\textrm{fm}$ was used in the plot.
For the case of UCOM contribution we have used the correlated
wave functions and the approximation $r-R_{-}(r) \approx R_{+}(r)-r$
for illustrative purpose \cite{radial}. Thus, the UCOM plot should
be taken only as a schematic one. From the figure one can see that
the Jastrow correlations cut out a significant part of the matrix
element at small $r$. This leads 
to a situation, where the total integrated areas under the radial
curve almost cancel out. 
In the case of UCOM correlations the cancellation is not as severe 
due to the fact that not so much amplitude is lost for small $r$. 

\begin{figure}[htb]
\includegraphics[width=12cm]{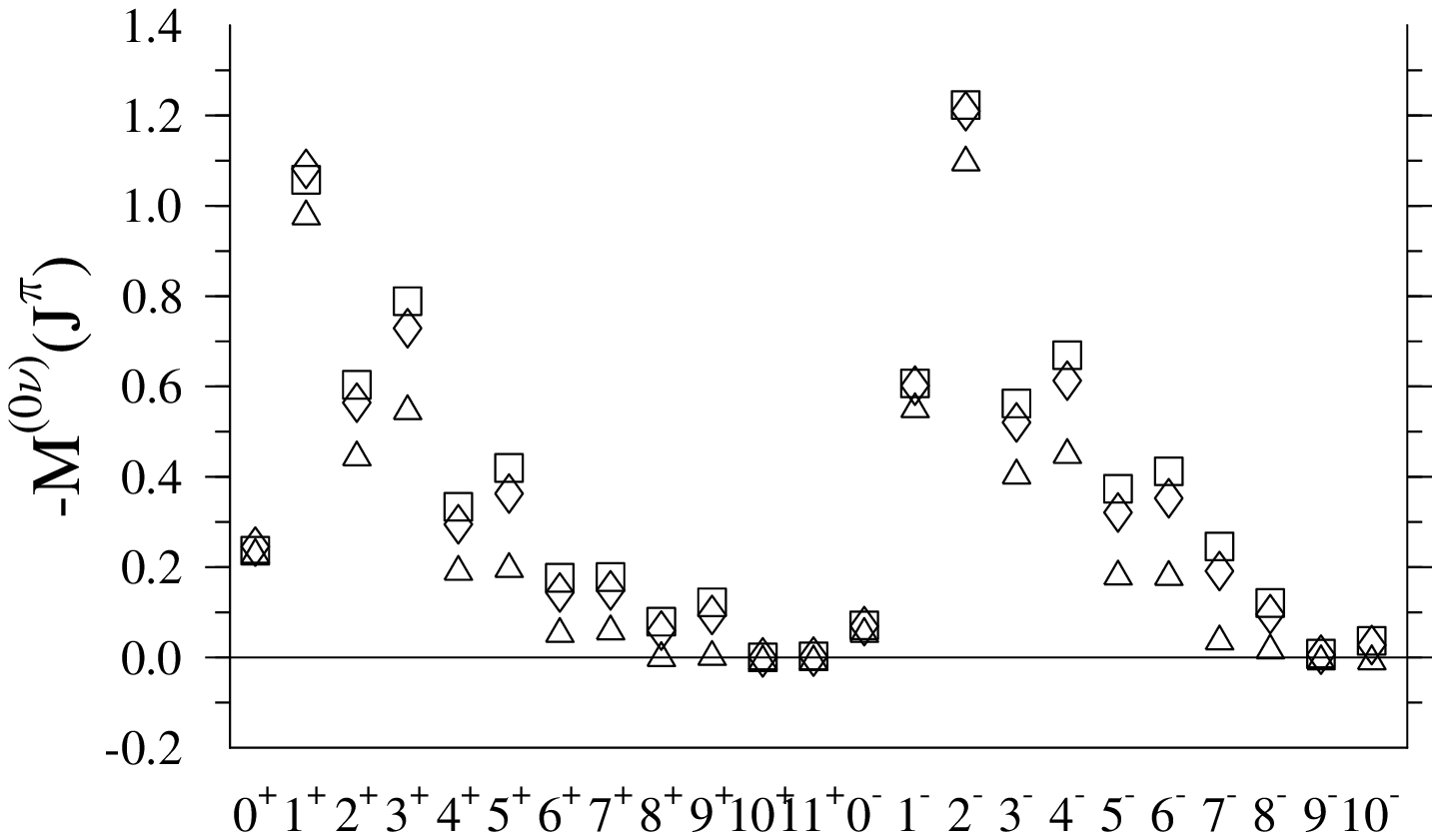}
\caption{The same as Fig.~\ref{f:totcomp} for the decay of $^{76}$Ge
  calculated by using the pnQRPA. Only the Bonn-A parametrization has
  been used for UCOM.}
\label{f:decom}
\end{figure}

Our results for the $0\nu\beta\beta$ decay of $^{76}$Ge are summarized
in Fig.~\ref{f:decom} and Table~\ref{t:qrpa}. The results have been
obtained by using the framework of the proton-neutron quasiparticle 
random-phase approximation (pnQRPA) \cite{teo2,JSbook}. The related
calculations, including the BCS and the pnQRPA calculations for
$^{76}$Ge and $^{76}$Se, were done in the model space consisting of
the single-particle 1p-0f-2s-1d-0g-0h$_{11/2}$ orbitals, both for
protons and neutrons. The single-particle energies were obtained from
a spherical Woods--Saxon potential. Slight adjustment was done for some of the
energies at the vicinity of the proton and neutron Fermi
surfaces to reproduce better the low-energy spectra of the neighboring
odd-$A$ nuclei and the low-energy spectrum of $^{76}$As.
The Bonn-A G-matrix \cite{HOL81} was used as a
two-body interaction and it was renormalized in the standard way, as
discussed e.g. in Refs. \cite{SUH88,SUH93,SUH02}. Due to this
phenomenological renormalization we did not perform an additional
UCOM renormalization of the two-body interaction. In the present
calculations we have used the `default value' $g_{\rm pp}=1.0$ for the
particle-particle interaction parameter of the pnQRPA.

In Fig.~\ref{f:decom} we display for $^{76}$Ge decay the multipole 
decomposition of the total $0\nu\beta\beta$ matrix element 
$M_{\rm GT}^{(0\nu)}-\left(g_{\rm V}/g_{\rm A}\right)^2M_{\rm F}^{(0\nu)}$
as derived from the pnQRPA calculations. The used symbols are the
same as in Fig.~\ref{f:totcomp}. The ratio $g_{\rm V}/g_{\rm A}=-1.254$
was used in the calculations. Since the nuclear wave functions have
been calculated by the use of the Bonn potential, we have used only
the Bonn-A parametrization for the UCOM.
Here one can see a pattern similar to
the case of $^{48}$Ca: the effect of the Jastrow
correlations grows strongly with increasing value of the angular momentum of
the intermediate states. As in the case of the $^{48}$Ca decay the
effect is the largest for the unnatural-parity states
$1^+,2^-,3^+,4^-,\ldots$ in an odd-odd nucleus. Contrary to the 
Jastrow-corrected multipole contributions,
the UCOM-corrected ones stay close to the bare contributions
for all intermediate multipoles $J^{\pi}$.

We summarize our results on the $0\nu\beta\beta$ matrix elements 
of the $^{76}$Ge decay in
Table~\ref{t:qrpa}, where we give the bare, Jastrow-corrected and
UCOM-corrected Gamow--Teller, Fermi and total matrix elements. 
For the total matrix element the Jastrow corrections
amount to 30\% reduction from the bare matrix element, whereas the
UCOM corrections are some 7\%. This, again, suggests that in
the earlier calculations \cite{teo1,tueb} the effect of the
short-range correlations has been considerably overestimated.

\begin{table}[h]
\caption{Gamow-Teller ($M_{\rm GT}^{(0\nu)}$),
Fermi ($M_{\rm F}^{(0\nu)}$) and total matrix elements derived from pnQRPA
calculations for the $0\nu\beta\beta$ decay of $^{76}$Ge.
The cases `bare', `Jastrow' and `UCOM' are as in Table \ref{t:gtcon}.
Only the Bonn-A parametrization has been used for UCOM.}
\label{t:qrpa}
\begin{center}\begin{tabular}{lrrc}
  & bare & Jastrow & UCOM Bonn-A \\
\hline
$M_{\rm GT}^{(0\nu)}$ & -6.755 & -4.681 & -6.265 \\
$M_{\rm F}^{(0\nu)}$  &  2.474 &  1.778 &  2.310 \\
total        & -8.328 & -5.811 & -7.734 \\
\end{tabular}\end{center}
\end{table}

In this Letter we have addressed the important issue of short-range
correlations in the context of neutrinoless double beta decay.
We have calculated the related nuclear matrix by the nuclear shell 
model for $^{48}$Ca and by the pnQRPA for $^{76}$Ge. The short-range
correlations have been calculated by the use of the simple Jastrow
function and the more refined UCOM method. Our computed results
indicate that the Jastrow method cuts off relevant parts of the many-body
wave function for high values of angular momentum of the
intermediate states. This leads to the excessive reduction of 30\% --
40\% in the magnitudes of the nuclear matrix elements. At the same
time the UCOM reduces the magnitudes of the matrix elements only by
7\% -- 16\%, roughly equally for all multipoles. Our results put to
question the recent calculations where short-range and tensor
correlations cause large effects on the nuclear matrix elements of 
neutrinoless double beta decay \cite{tueb}.
Study of the effects of the UCOM procedure upon heavier nuclei is in
progress.

{\bf Acknowledgements:} This work has been partially supported by the 
National Research Council (CONICET) of Argentina and by the
Academy of Finland under the Finnish Centre of Excellence Programme 
2006-2011 (Nuclear and Accelerator Based Programme at JYFL). We thank
also the EU ILIAS project under the contract RII3-CT-2004-506222.
One of the authors (O.C.) gratefully thanks for the warm hospitality 
extended to him at the Department of Physics of the University of 
Jyv\"askyl\"a, Finland.


\begin{thebibliography}{99}
\bibitem{exp1} V. Tretyak, Y. Zdesenko, At. Data Nucl. Data Tables
80 (2002) 83.
\bibitem{compil} S.R. Elliott, P. Vogel, Ann. Rev. Nucl. Part.
Sci. 52 (2002) 115.
\bibitem{teo1} T. Tomoda, Rep. Prog. Phys. 54 (1991) 53.
\bibitem{teo2} J. Suhonen, O. Civitarese, Phys. Rep. 300
    (1998) 123.
\bibitem{DOI85} M. Doi, T. Kotani, E. Takasugi, Prog.
Theor. Phys. Suppl. 83 (1985) 1.
\bibitem{miller} G.A. Miller, J.E. Spencer, Ann. Phys. 100 (1976) 562.
\bibitem{tueb} V.A. Rodin, A. Faessler, F. \v Simkovic, P. Vogel,
Nucl. Phys. A 766 (2006) 107.
\bibitem{KRM92} F. Krmpotic, J. Hirsch, H. Dias,
Nucl. Phys. A 542 (1992) 85.
\bibitem{BRO77} G.E. Brown, S.O. B\"ackman, E. Oset, W. Weise,
Nucl. Phys. A 286 (1977) 191.
\bibitem{TOW87} I.S. Towner, Phys. Rep. 155 (1987) 263.
\bibitem{BAR99} C. Barbero, F. Krmpotic, A. Mariano, D. Tadic,
Nucl. Phys. A 650 (1999) 485.
\bibitem{feldmeier98} H. Feldmeier, T. Neff, R. Roth, J. Schnack,
Nucl. Phys. A 632 (1998) 61.
\bibitem{roth05} R. Roth, H. Hergert, P. Papakonstantinou, T. Neff,
H. Feldmeier, Phys. Rev. C 72 (2005) 034002.
\bibitem{shell} B.A. Brown, A. Etchegoyen, W.D.M. Rae,
The computer code OXBASH, MSU-NSCL Report 524 (1988).
\bibitem{force} W.A. Richter, M.G. Van Der Merwe, R.E. Julies,
B.A. Brown, Nucl. Phys. A 523 (1991) 325.
\bibitem{neff03} T. Neff, H. Feldmeier, Nucl. Phys. A 713 
(2003) 311.
\bibitem{RET95} J. Retamosa, E. Caurier, F. Nowacki, Phys. Rev.
 C 51 (1995) 371.
\bibitem{radial} In all other numerical applications of the UCOM 
we have used correlated operators without involving any
approximations. 
\bibitem{JSbook} J. Suhonen, From Nucleons to Nucleus: Concepts
  of Microscopic Nuclear Theory, Springer, Berlin, 2007.
\bibitem{HOL81} K. Holinde, Phys. Rep. 68 (1981) 121.
\bibitem{SUH88} J. Suhonen, T. Taigel, A. Faessler,
Nucl. Phys. A 486 (1988) 91.
\bibitem{SUH93} J. Suhonen, Nucl. Phys. A 563 (1993) 205.
\bibitem{SUH02} J. Suhonen, Nucl. Phys. A 700 (2002) 649.
\end{thebibliography}
\end{document}